# OGBoost: A **Python** Package for Ordinal Gradient Boosting

**Mansour T.A. Sharabiani** 　**Alex Bottle** 　**Alireza S. Mahani**
Imperial College London　Imperial College London　Statman Solution Ltd.

## Abstract

This paper introduces **OGBoost**, a **scikit-learn**-compatible `Python` package for ordinal regression using gradient boosting. Ordinal variables (e.g., rating scales, quality assessments) lie between nominal and continuous data, necessitating specialized methods that reflect their inherent ordering. Built on a coordinate-descent approach for optimization and the latent-variable framework for ordinal regression, **OGBoost** performs joint optimization of a latent continuous regression function (functional gradient descent) and a threshold vector that converts the latent continuous value into discrete class probabilities (classical gradient descent). In addition to the stanadard methods for **scikit-learn** classifiers, the `GradientBoostingOrdinal` class implements a `decision_function` that returns the (scalar) value of the latent function for each observation, which can be used as a high-resolution alternative to class labels for comparing and ranking observations. The class has the option to use cross-validation for early stopping rather than a single holdout validation set, a more robust approach for small and/or imbalanced datasets. Furthermore, users can select base learners with different underlying algorithms and/or hyperparameters for use throughout the boosting iterations, resulting in a 'heterogeneous' ensemble approach that can be used as a more efficient alternative to hyperparameter tuning (e.g. via grid search). We illustrate the capabilities of **OGBoost** through examples, using the wine quality dataset from the UCI respository. The package is available on PyPI and can be installed via `pip install ogboost`.

## 1. Introduction

Ordinal response variables frequently arise in applied research, such as rating surveys (e.g., "strongly disagree" to "strongly agree"), product quality assessments (e.g., "poor" to "excellent"), and medical scoring systems (e.g., severity of postoperative kidney injuries (Khwaja 2012)). These outcomes possess a natural ordering but lack precise numerical distances between categories. Consequently, ordinal data require specialized methods that account for their unique properties.

In the `Python` ecosystem, existing tools for handling ordinal data have significant limitations. For instance, ordered logistic and probit regression models, implemented in packages like **statsmodels**, assume a linear relationship between features and the mean of a cumulative density function that is combined with a set of thresholds to generate the discrete class labels. These linear ordinal regression models are too rigid to capture complex, nonlinear effects of features on the outcome. Machine learning classifiers, such as those in the **scikit-learn** Python package, treat ordinal outcomes as nominal, discarding the valuable ordering information. Regressors, which treat ordinal outcomes as continuous, implicitly assume equidistant categories and/or unbounded outcomes, both unrealistic assumptions. Ranking models, such as those available in **LightGBM** (Ke, Meng, Finley, Wang, Chen, Ma, Ye, and Liu 2017) and **XGBoost** (Chen and Guestrin 2016), focus on pairwise or listwise comparisons, which are useful for ranking tasks but do not directly address ordinal regression, where absolute categories are of primary importance. The absence of a dedicated abstraction for ordinal outcomes in **scikit-learn**, similar to the well-defined interfaces for classifiers and regressors, further highlights the gap.

This paper introduces **OGBoost** (which stands for 'Ordinal Gradient Boosting'), a Python package that enables the use of any machine learning regression model - including decision trees, neural networks and support vector machines - for ordinal outcomes through a generalization of the standard gradient boosting algorithm. Following a coordinate-descent strategy (Wright 2015), **OGBoost** alternates between 1) updating a latent regression function, $g(\mathbf{x})$, which represents the mean of a probability density function (PDF), and 2) refining a threshold vector $\theta$, which produces a finite set of probabilities - each associated with an ordinal category - from the PDF. It may be helpful to think of coordinate-descent as the optimization counterpart to Gibbs sampling, including the 'Multivariate-from-Univariate' (MfU) Markov Chain Monte Carlo sampler (Mahani and Sharabiani 2017).

Previous research on machine learning approaches for ordinal regression goes back to the early 2000s and spans several methodologies. Early efforts focused on adapting support vector machines to handle ordered outcomes. For example, (Chu and Keerthi 2005) extended the SVM framework by incorporating threshold constraints, thereby ensuring that the predicted labels respected the natural ordering of the classes. Subsequent work explored kernel methods to further capture the nonlinear relationships inherent in ordinal data (Sun, Li, Wu, Zhang, and Li 2009). More recently, deep learning has been applied to ordinal tasks in areas such as age estimation (Xie and Pun 2020) and medical diagnosis (Liu, Zou, Song, Yang, You, and K Vijaya Kumar 2018). Despite these varied approaches, most methods remain tied to specific model classes or require complex modifications to conventional algorithms, and have not resulted in general-purpose, open-source software. In contrast, **OGBoost** provides a unified, **scikit-learn**-compatible framework that allows any machine learning regression model to be seamlessly adapted for ordinal regression through gradient boosting and coordinate descent. This general approach facilitates systematic exploration of complex, nonlinear effects in ordinal data while preserving the intrinsic ordering of the response.

The closest software counterpart to **OGBoost** is the Python package **OrdinalGBT** (which stands for 'Ordinal Gradient Boosted Trees'). Despite their common goal and similarities, **OGBoost** offers several advantages over **OrdinalGBT**. First, **OrdinalGBT** is limited to using decision trees as base learner, whereas **OGBoost** allows users to choose any ML regression model as base learner. Even further, **OGBoost** allows users to use different base learners at each boosting iteration, resulting in 'heterogeneous boosting ensembles'. Secondly, **OrdinalGBT** outsources the gradient boosting algorithm to **LightGBM** in order to take advantage of its speed and scalability. While this design choice can be beneficial for large datasets, it also prevents **OrdinalGBT** from performing a genuine, joint optimization of the regression function and the threshold vector. Instead, the thresholds in **OrdinalGBT** are set before the gradient boosting process begins. In contrast, the coordinate-descent approach of **OGBoost** alternates between improving the regression function and thresholds at each boosting iteration, thus leading to a more systematic exploration of the model parameter space.

The remainder of this paper is organized as follows. We first introduce the mathematical framework underlying **OGBoost**. This is followed by an overview of the components and key features of **OGBoost**. We then demonstrate the practical application of the package using the Wine Quality Dataset, including the use of latent-function prediction for ranking, early-stopping using cross-validation, comparison with linear ordinal models, and heterogeneous ensembles as an alternative to hyperparameter tuning. Finally, we conclude with a discussion of future directions for research and development.

## 2. Mathematical Framework

We adopt the threshold (or cumulative link) model (Gutiérrez, Perez-Ortiz, Sanchez-Monedero, Fernandez-Navarro, and Hervas-Martinez 2015) for generating ordinal responses $y \in \{0, 1, \ldots, M-1\}$. In this framework, each observation is associated with a latent variable generated as

$$z(\mathbf{x}) = g(\mathbf{x}) + \epsilon,$$

where $g(\mathbf{x})$ is a (potentially nonlinear) function of the predictors $\mathbf{x}$ that shifts the mean of the distribution, and $\epsilon$ is a random deviate drawn from a continuous distribution with zero mean and fixed dispersion. The distribution is characterized by a probability density function (PDF) $\phi(\cdot)$ and a cumulative distribution function (CDF) $\Phi(\cdot)$.

We introduce a vector of thresholds

$$\boldsymbol{\theta} = (\theta_0, \theta_1, \ldots, \theta_M),$$

where the endpoints are fixed as

$$\theta_0 = -\infty \quad \text{and} \quad \theta_M = +\infty.$$

These thresholds partition the real line into $M$ intervals. The cumulative link model posits that the probability that the response is less than or equal to a given ordinal category $m$ is

$$P(y \leq m \mid \mathbf{x}) = \Phi(\theta_{m+1} - g(\mathbf{x})).$$

Thus, the probability of observing class $m$ is given by the difference

$$P(y = m \mid \mathbf{x}) = \Phi(\theta_{m+1} - g(\mathbf{x})) - \Phi(\theta_m - g(\mathbf{x})).$$

For example, if $\epsilon$ follows a standard normal distribution, then $\Phi$ is the standard normal CDF (yielding an ordered probit model); if $\epsilon$ follows a logistic distribution, the logistic CDF is used (yielding an ordered logit model).

The negative log-likelihood for $N$ observations is therefore expressed as

$$\mathcal{L}(g, \boldsymbol{\theta}) = -\sum_{n=1}^{N} \ln\left[\Phi(\theta_{y_n+1} - g(\mathbf{x}_n)) - \Phi(\theta_{y_n} - g(\mathbf{x}_n))\right].$$

Our procedure minimizes this loss via a coordinate descent algorithm that alternates between:

1. **Updating $g(\mathbf{x})$ via gradient boosting:** For each observation $n$, we compute the pseudo-residual $-\frac{\partial \mathcal{L}}{\partial g(\mathbf{x}_n)}$, fit a base learner (e.g., a regression tree) to these residuals, and update $g(\mathbf{x})$ using a shrunken version of the new learner's predictions.

2. **Updating the thresholds $\boldsymbol{\theta}$:** We compute the partial derivatives $\frac{\partial \mathcal{L}}{\partial \theta_m}$ for $m = 1, \ldots, M-1$ and perform a line search that adjusts the thresholds while maintaining their strict ordering and reducing the loss.

## 2.1. Derivative with Respect to $g(\mathbf{x})$

At each observation $(\mathbf{x}_n, y_n)$, the derivative of the per-observation loss $l_n$ with respect to $g(\mathbf{x}_n)$ boils down to the difference of PDFs over the cumulative difference in the denominator:

$$\frac{\partial l_n}{\partial g(\mathbf{x}_n)} = -\frac{\phi(\theta_{y_n} - g(\mathbf{x}_n)) - \phi(\theta_{y_n+1} - g(\mathbf{x}_n))}{\Phi(\theta_{y_n+1} - g(\mathbf{x}_n)) - \Phi(\theta_{y_n} - g(\mathbf{x}_n))},$$

where $\phi(\cdot)$ is the standard normal PDF. The negative of this derivative serves as the *pseudo-residual* for gradient boosting. See Appendix A for details.

## 2.2. Derivative with Respect to $\theta_m$

The derivative of $\mathcal{L}$ w.r.t. each threshold $\theta_m$ is obtained by focusing on how $\theta_m$ affects the upper/lower boundaries of the integrals in the likelihood terms:

$$\frac{\partial \mathcal{L}}{\partial \theta_m} = -\sum_{n=1}^{N} \frac{\phi(\theta_m - g(\mathbf{x}_n)) \left[\delta_{y_n, m} - \delta_{y_n, m+1}\right]}{\Phi(\theta_{y_n+1} - g(\mathbf{x}_n)) - \Phi(\theta_{y_n} - g(\mathbf{x}_n))},$$

where $\delta_{i,j}$ is the Kronecker delta. An equivalent formulation groups observations by $y_n$, yielding a direct sum over sets of indices with $y_n = m$ or $y_n = m + 1$. See Appendix A for details.

## 2.3. Initialization of g and theta

We initialize the latent function $g_0(\mathbf{x})$ to a constant (often zero). This simplifies finding an initial $\theta$ by matching empirical class frequencies. Specifically, if $g(\mathbf{x}_n) \equiv 0$, then the frequency of class $m$ should approximate

$$\Phi(\theta_{m+1}) - \Phi(\theta_m),$$

and solving these equations produces threshold estimates $\theta_m = \Phi^{-1}\left(\sum_{j=0}^{m} \hat{p}_j\right)$, where $\hat{p}_m$ is the empirical proportion of class $m$. See Appendix B.

## 2.4. Learning Rate and Line Search

Each update to $g(\mathbf{x})$ is scaled by a *learning rate*, $\eta_g$, which is typically a small fraction (e.g. 0.1). Meanwhile, threshold updates can use a separate rate, $\eta_\theta$, which is often adaptively tuned using a doubling/halving line search. We accept or reject a proposed update $\Delta \theta = -\eta_\theta \nabla_\theta \mathcal{L}$ based on:

$$\mathcal{L}(\theta - \Delta \theta) < \mathcal{L}(\theta), \quad \text{and} \quad \theta_m - \Delta \theta_m < \theta_{m+1} - \Delta \theta_{m+1} \;\; \forall m,$$

which ensures that thresholds remain strictly ordered. See Appendix C for full pseudo-code and additional details.

# 3. Package Overview

**OGBoost** consists of two classes (`GradientBoostingOrdinal` and `LinkFunction`), and three functions (`concordance_index`, `generate_heterogeneous_learners`, and `load_wine_quality`). An overview of these components is provided in Table 1.

Table 1: Overview of **OGBoost** Components

| Component | Description |
|---|---|
| `GradientBoostingOrdinal` | The main class, a **scikit-learn**-compatible estimator that extends the classifier interface to support ordinal regression. |
| `LinkFunction` | Provides a common interface for transforming latent scores into probabilities. <br><br> • Supported transformations: probit, logit, and cloglog. |
| `concordance_index` | Evaluates model performance for ordinal data by computing the proportion of pairs where the predicted ordering matches the true ordering. |
| `generate_heterogeneous_learners` | Creates a list of base learners with different hyperparameters, allowing the training of a heterogeneous boosting ensemble model. |
| `load_wine_quality` | Loads the Wine Quality datasets (red and white) from the UCI ML repository (or from package cache). |

**OGBoost** is fully compatible with **scikit-learn** and can be used in the same way as any other **scikit-learn** estimator. This includes support for hyperparameter tuning using `GridSearchCV` and `RandomizedSearchCV`, integration with other **scikit-learn** tools such as pipelines and transformers, and compatibility with MLOps frameworks such as **MLflow** (Chen, Chow, Davidson, DCunha, Ghodsi, Hong, Konwinski, Mewald, Murching, Nykodym *et al.* 2020).

To maximize ease of use for those already familiar with **scikit-learn**, `GradientBoostingOrdinal` follows the same naming conventions for its class initializers (e.g., `n_estimators`, `learning_rate`, `n_iter_no_change`) and methods (e.g., `predict`/`predict_proba`/`decision_function` and their `staged_...` counterparts) as **scikit-learn**'s `GradientBoostingClassifier` class.

## 3.1. Key Features

While **OGBoost** follows the **scikit-learn** API conventions, it also introduces several novel features that are tailored to the unique requirements of ordinal regression. We discuss them below.

*Latent Score Prediction*

In `GradientBoostingOrdinal`, the `decision_function` method returns the value of the latent function $g(\mathbf{x})$ for each observation. The concordance index, which measures the proportion of pairs where the order of predictions and true labels is the same, is typically higher when using the latent scores instead of class labels. This is because the latent scores retain a higher resolution than the discrete class labels, allowing for more accurate ranking of observations. See Section 4.3 for an example.

Note that in **scikit-learn** classifiers, the `decision_function` method behaves differently as it returns not a single value but a vector of values (one for each class) representing the distance to the decision boundaries. Therefore, when using nominal classifiers for ordinal regression, one can only use the discrete class labels for ranking.

### *The* `score` *Method*

The default behavior of the `score` method in `GradientBoostingOrdinal` is to use the concordance index as metric - instead of accuracy, and the latent score predictions - rather than class labels. The former is more appropriate for ordinal regression, while the latter retains more information than the discrete class labels.

### *Heterogeneous Ensembles*

While nearly all software implementations of gradient boosting exclusively use decision trees, **OGBoost** allows users to choose any regression model as the base learner. Even further, users can provide a list - or a generator - of base learners to the `GradientBoostingOrdinal` initializer, resulting in a 'heterogeneous' boosting ensemble. These can be different learning algorithms (e.g. decision trees and neural networks), or a single algorithm with different hyperparameters (e.g. decision trees with different maximum depth). Such built-in diversity can reduce the need for explicit model selection and hyperparameter tuning, thus saving significant computational resources, while improving model performance by leveraging the strengths of different algorithms. See Section 4.6 for an example.

### *Cross-Validation for Early Stopping*

In addition to using a single validation set for early stopping, `GradientBoostingOrdinal` can use cross-validation for this purpose. This feature is particularly useful for small and/or imbalanced datasets, where a small validation set may not be representative of the overall data distribution. By using cross-validation, **OGBoost** the generalization performance of the model as a function of boosting iteration can be estimated more robustly, leading to better performance in unseen data. See Section 4.5 for an illustration.

## 4. Using OGBoost

Assuming that readers are familiar with the **scikit-learn** API and its conventions for model training, prediction, and evaluation, we skip such details for **OGBoost**. Insead, the examples in this section highlight a few distinct features of **OGBoost**. More specifically, we cover these topics:

1. Plotting training and validation loss for diagnostics,

2. Discriminative performance of discrete class labels vs. continuous latent scores,

3. Performance comparison with linear ordinal regression,

4. Cross-validation vs. single validation set for early stopping, and

5. Heterogeneous boosting ensembles vs. hyperparameter tuning.

### 4.1. Data Loading and Preparation

The Wine Quality Dataset is a well-known benchmark for evaluating machine learning algorithms. It contains information about various physicochemical properties of red and white wines, along with their quality ratings on a nominal scale from low to high. The dataset is available from the UCI Machine Learning Repository and can be easily loaded using the `load_wine_quality` function in **OGBoost**. This function downloads the dataset if it is not already cached locally, and adjusts the target variable to start from 0, matching **OGBoost**'s convention.

The code block below shows how to load the dataset (the red portion) and split it into training and test sets. We use a stratified split to ensure that the class distribution is preserved between the training and test sets. We also set a random seed for reproducibility.

```python
from ogboost import load_wine_quality
from sklearn.model_selection import train_test_split
random_seed = 123
X, y, _, _ = load_wine_quality(return_X_y=True)
X_train, X_test, y_train, y_test = train_test_split(
    X, y, test_size=0.3,
    stratify=y,
    random_state=random_seed
)
```

### 4.2. Training and Validation Loss

`GradientBoostingOrdinal` includes a `plot_loss` method that visualizes the training and validation loss over the boosting iterations. This is useful for diagnosing the model's performance and checking for overfitting. It also shows the loss improvement at each iteration, for regression-function updates and threshold updates separately. This is helpful for understanding the contribution of each update to the overall loss improvement, including whether both are indeed improving the model.

```python
from ogboost import GradientBoostingOrdinal
ogb = GradientBoostingOrdinal(
  n_iter_no_change=10,
  random_state=random_seed
).fit(X_train, y_train)
ogb.plot_loss()
```

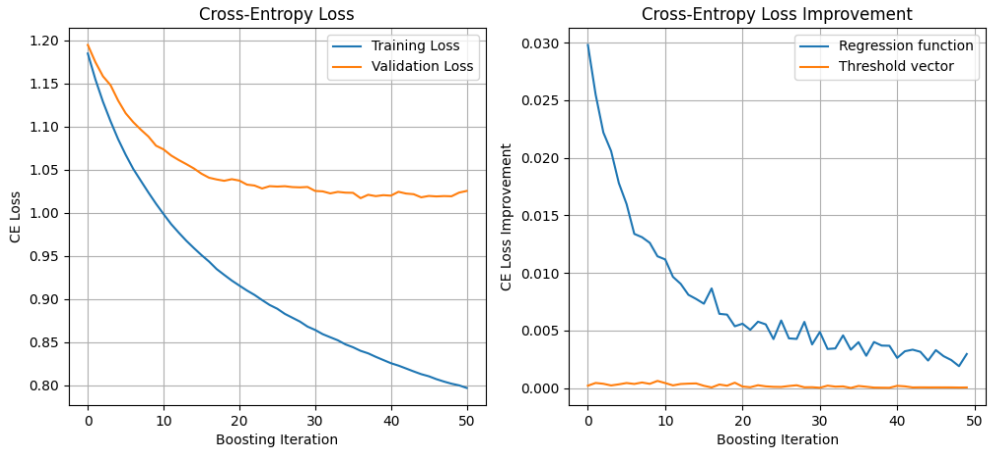

The plot can be customized by passing keyword arguments to the `plot_loss` method, such as `figsize`, `training_style`, `validation_style`, etc. See method documentation for more details.

### 4.3. Discrete Class Labels vs. Continuous Latent Score

An important advantage of ordinal regression over nominal classification is the ability to use the predicted latent function $g(\mathbf{x})$ - rather than the discrete class labels - for ranking observations. This can be done using the `decision_function` method:

```
cindex_latent_scores = ogb.score(X_test, y_test)
cindex_class_labels = ogb.score(X_test, y_test, pred_type='labels')
print(f'Concordance (class labels): {cindex_class_labels:.3f}')
print(f'Concordance (latent scores): {cindex_latent_scores:.3f}')
```

==================================================

Concordance (class labels): 0.723

Concordance (latent scores): 0.794

==================================================

The concordance index for the latent scores is significantly higher than class labels (79.4% vs. 72.3%), indicating that the latent scores rank observations more accurately than the class labels. This is expected, as the discrete class labels lose information when converted from the continuous latent scores.

### 4.4. Linear Ordinal Regression

It is helpful to compare the performance of **OGBoost** against a linear ordinal regression model. We use the `OrderedModel` class from **statsmodels** to fit a linear ordered-probit model to the training data. To assess the statistical significance of the difference in performance between the two models, we use the fold-level numbers from repeated K-fold cross-validation to perform a paired t-test.

In order to utilize **scikit-learn**'s utility function `cross_val_score`, we wrap `OrderedModel` in a custom class, `OrderedModelWrapper`, that implements the minimum required methods. The source code for `OrderedModelWrapper` is available in the `replication_library` module (see Supplementary Material).

```python
from sklearn.model_selection import RepeatedKFold
n_splits = 10
n_repeats = 10
rkf = RepeatedKFold(
  n_splits=n_splits,
  n_repeats=n_repeats,
  random_state=random_seed
)

from sklearn.model_selection import cross_val_score
from replication_material_library import OrderedModelWrapper
cv_linear = cross_val_score(
    OrderedModelWrapper(),
    X, y, cv=rkf,
    n_jobs=-1
)
cv_ogboost = cross_val_score(
    GradientBoostingOrdinal(random_state=random_seed),
    X, y, cv=rkf,
    n_jobs=-1
)
print(f'Concordance (linear ordinal regression): {cv_linear.mean():.3f}')
print(f'Concordance (OGBoost): {cv_ogboost.mean():.3f}')
t_stat, p_value = ttest_rel(cv_ogboost, cv_linear)
print(f'T-statistic: {t_stat:.3f}, p-value: {p_value:.3f}')
```

==================================================

Concordance (linear ordinal regression): 0.796

Concordance (OGBoost): 0.822

T-statistic: 15.559, p-value: 0.000

==================================================

We see that **OGBoost** outperforms the linear ordinal regression model with statistical significance. This is expected, as **OGBoost** can capture nonlinear relationships between predictors and outcomes, unlike the linear ordinal regression model.

### 4.5. Early Stopping with Cross-Validation

In small or imbalanced datasets, a single validation set - which is typically a small fraction of the overall data - may not be representative of the overall data distribution. In such cases, cross-validated loss can provide a more robust estimate of the model's generalization performance as the boosting algorithm progresses, and thus lead to a more optimal early stopping decision. **OGBoost** offers this feature via the `cv_early_stopping_splits` parameter. (Note that the final model used for prediction is a single model trained on the entire training set, and the CV splits are used only for early stopping.)

The code snippet below compares the two approaches using repeated K-fold cross-validation and a statistical test to determine if the difference in performance is significant:

```python
earlystop_simple = GradientBoostingOrdinal(
  n_iter_no_change=10,
  validation_fraction=0.2,
  random_state=random_seed
)
cv_earlystop_simple = cross_val_score(earlystop_simple, X, y, cv=rkf, n_jobs=-1)
earlystop_cv = GradientBoostingOrdinal(
  n_iter_no_change=10,
  cv_early_stopping_splits=5,
  random_state=random_seed
)
cv_earlystop_cv = cross_val_score(earlystop_cv, X, y, cv=rkf, n_jobs=-1)
print(f'Concordance (simple early stopping): {cv_earlystop_simple.mean():.3f}')
print(f'Concordance (CV early stopping): {cv_earlystop_cv.mean():.3f}')
t_stat, p_value = ttest_rel(cv_earlystop_cv, cv_earlystop_simple)
print(f'T-statistic: {t_stat:.3f}, p-value: {p_value:.3f}')
```

==================================================

Concordance (simple early stopping): 0.818

Concordance (CV early stopping): 0.820

T-statistic: 2.362, p-value: 0.020

==================================================

The cross-validated early stopping approach leads to a slightly higher - though statistically significant - concordance index than the simple early stopping approach.

### 4.6. Heterogeneous Boosting Ensembles

A useful way to motivate the heterogenous ensemble method for gradient boosting is by comparing it to the standard ML approach for hyperparameter tuning, i.e., grid search.

*Grid Search*

Assume we want to tune the `max_depth` parameter of the `DecisionTreeRegressor` base learner in `GradientBoostingOrdinal`, using the following candidate list:

```python
candidates_dt_max_depth = [3, 6, 9, None]
```

We can set up a `GridSearchCV` learner from the above list. To identify the correct parameter name for defining the parameter grid, we use the `get_params` method:

```python
from sklearn.tree import DecisionTreeRegressor
from ogboost import GradientBoostingOrdinal
GradientBoostingOrdinal(base_learner=DecisionTreeRegressor()).get_params()
```

Inspecting the output reveals that the desired parameter name is `base_learner__max_depth`. We can now set up the grid search learner and benchmark it using repeated K-fold cross-validation:

```python
from sklearn.model_selection import GridSearchCV
from sklearn.tree import DecisionTreeRegressor
import time
dt_grid = {
    'base_learner__max_depth': candidates_dt_max_depth,

}
n_samples = 100
kf_gridsearch = KFold(n_splits=5, shuffle=True, random_state=random_seed)
dt_learner_gridsearch = GridSearchCV(
    GradientBoostingOrdinal(
        base_learner=DecisionTreeRegressor(),
        n_estimators=n_samples,
        random_state=random_seed
    ),
    param_grid=dt_grid,
    cv=kf_gridsearch
)
start = time.time()
cv_gridsearch = cross_val_score(dt_learner_gridsearch, X, y, cv=rkf, n_jobs=-1)
end = time.time()
print(f'Concordance (GridSearchCV): {cv_gridsearch.mean():.3f}')
print(f'Time: {t_gridsearch:.1f} seconds')
```

======================================================

Concordance (GridSearchCV): 0.849

Time: 780.3 seconds

======================================================

The `GridSearchCV` learner requires training and prediction for all folds (5 here) and all hyperparameter values in the grid (4 here), leading to high training count (5x4=20 here) and thus long training times.

*Heterogeneous Ensemble*

Alternatively, we can change our perspective from *selecting* (the best hyperparameter) to *combining* (all hyperparameters). The latter is what we refer to as a *heterogenous ensemble* in the context of gradient boosting. The core idea is to randomly select a hyperparameter from the candidate list in each boosting iteration and form an ensemble of decision trees with different maximum-depth parameters, letting the gradient boosting algorithm choose the weight of each learner automatically.

All is needed is to define a lambda for generating samples from the candidate list:

```python
dt_overrides = {
    "max_depth": lambda rng: rng.choice(candidates_dt_max_depth),
}
```

Note that in the above, we do not need to add the `base_learner` prefix to the parameter name. We can now use the `generate_heterogeneous_learners` utility function provided in the **OGBoost** package to randomly generate a list of base learners with different hyperparameters, and pass it to `GradientBoostingOrdinal`:

```python
from ogboost import generate_heterogeneous_learners
dt_template = DecisionTreeRegressor()
dt_overrides = {
    "max_depth": lambda rng: rng.choice(candidates_dt_max_depth),
}
random_learners = generate_heterogeneous_learners(
    [dt_template], [dt_overrides],
    total_samples=n_samples,
    random_state=random_seed,
)
ogboost_hetero = GradientBoostingOrdinal(
    base_learner=random_learners,
    n_estimators=n_samples,
    random_state=random_seed
)
```

We can benchmark this learner using the same repeated k-fold object as before:

```python
start = time.time()
cv_hetero = cross_val_score(ogboost_hetero, X, y, cv=rkf, n_jobs=-1)
end = time.time()
print(f'Concordance (Heterogeneous ensemble): {cv_hetero.mean():.3f}')
print(f'Time: {t_hetero:.1f} seconds')
```

==================================================

Concordance (Heterogeneous ensemble): 0.861

Time: 46.0 seconds

==================================================

We see that, despite taking only a fraction of time (nearly 20x less), the heterogeneous ensemble has more than a full percentage point advantage over the grid search method.

Statistical testing confirms the significance of improvement:

```python
from scipy import stats
t_stat, p_value = stats.ttest_rel(cv_hetero, cv_gridsearch_dt)
print(f't-statistic: {t_stat:.3f}, p-value: {p_value:.3f}')
```

==================================================

T-statistic: 9.624, p-value: 0.000

==================================================

While these results are based on one example, and benchmarking studies using multiple datasets are needed, yet they suggest that heterogenous boosting ensembles can be an efficient alternative to hyperparameter tuning strategies, and their implementation in mainstream gradient boosting libraries such as **LighGBM** and **XGboost** - as well as their systematic testing - may be warranted.

Finally, note that **OGBoost** also supports heterogeneous ensembles of different algorithms, not just different hyperparameters of the same algorithm. This is achieved by passing a list of different base learners to `GradientBoostingOrdinal`:

```python
from sklearn.neural_network import MLPRegressor
mlp_template = MLPRegressor(max_iter=500)
mlp_overrides = {
    "hidden_layer_sizes": lambda rng: (rng.choice([10, 20, 30]),)
}
dt_mlp_learners = generate_heterogeneous_learners(
    [dt_template, mlp_template],
    [dt_overrides, mlp_overrides],
    total_samples=n_samples,
    random_state=random_seed,
    template_probs=[0.75, 0.25]
)
ogboost_hetero_v2 = GradientBoostingOrdinal(
    base_learner=dt_mlp_learners,
    n_estimators=n_samples,
    random_state=random_seed
).fit(X_train, y_train)
```

Note that, when specifying more than one base learner, the `template_probs` parameter is used to specify the probability of selecting each base learner. In the above example, we set the probability of selecting a decision tree to 0.75 and that of selecting a neural network to 0.25.

## 5. Discussion

In this article, we presented **OGBoost**, a **scikit-learn**-compatible Python package for ordinal regression using gradient boosting. By employing a coordinate-descent strategy, **OGBoost** jointly optimizes a latent regression function $g(\mathbf{x})$ and a threshold vector $\boldsymbol{\theta}$ (with fixed endpoints $\theta_0 = -\infty$ and $\theta_M = +\infty$) to model ordinal outcomes. Important features of **OGBoost** include continuous latent score as an alternative to discrete class labels for ranking tasks, option to use of any ML regression model as the base learner, cross-validation as a more robust alternative to a single validation set for early stopping, and heterogeneous ensembles of different models and hyperparameters as an alternative to model selection and hyperparameter tuning. We demonstrated these features of **OGBoost** on the Wine Quality Dataset.

The complete freedom in **OGBoost** to select any **scikit-learn** regressor - or any heterogeneous sequence of varying algorithms and/or hyperparameter configurations - is in stark contrast to many existing gradient boosting libraries, which restrict users to a single model instance, often a decision tree, as the base learner. However, by relying on existing **scikit-learn** regressor classes, the current **OGBoost** implementation does not incorporate advanced optimization techniques - such as those employed by LightGBM and XGBoost - to accelerate training and improve scalability.

For example, **LightGBM** and **XGBoost** utilize histogram-based feature discretization, gradient-based one-side sampling (GOSS), and exclusive feature bundling (EFB) to optimize tree-based learning on large-scale datasets. These strategies could be adapted to a specialized version of **OGBoost** focused on decision trees, while the current framework remains available for other choices of base learners.

Another area for future improvement is the cross-validation (CV)-based early stopping mechanism. Although CV-based early stopping enhances robustness, especially for small or imbalanced datasets, it adds additional computational overhead. Future work could explore parallelizing the training across CV folds. Alternatively, instead of retraining a single model on the entire dataset after early stopping, an aggregate of the fold-level models (e.g., via averaging or voting) could be used to form the final predictive model, potentially reducing training time and leveraging ensemble diversity.

The current implementation of **OGBoost** employs the negative log-likelihood loss, which is natural for probabilistic models. However, alternative loss functions might be advantageous in certain settings. For instance, ranking-based losses (such as a hinge loss adapted for ordinal data), robust losses like the Huber loss, or cost-sensitive losses that account for varying misclassification costs across ordinal classes could offer improvements in accuracy or interpretability. Exploring these alternatives and understanding their impact on performance is an important avenue for future research.

Beyond these algorithmic improvements, a comprehensive benchmarking effort is needed. Future work should systematically compare the heterogeneous boosting ensemble approach - both in terms of prediction quality and computational efficiency - against grid-search-based hyperparameter tuning, not only for ordinal regression but also for nominal classification and continuous regression tasks. In addition, evaluating **OGBoost** across diverse datasets, including those from healthcare, education, and social sciences, would provide valuable insights into its practical applicability and robustness.

In summary, while **OGBoost** already demonstrates significant improvements over existing methods, several promising directions remain for enhancing its efficiency and effectiveness. These include specialized optimization for decision-tree learners, parallelization of cross-validation, exploration of alternative loss functions tailored to ordinal data, and extensive cross-domain benchmarking. We believe that addressing these areas will further solidify the role of **OGBoost** as a versatile and powerful tool for ordinal regression.

# 6. Compute Environment

The following system specifications were captured at runtime:

- OS: Windows
- Kernel: 10
- OS Version: 10.0.26100
- Processor: Intel(R) Core(TM) 7 150U
- CPU Architecture: 64bit
- Physical Cores: 10
- Logical Cores: 12
- CPU Frequency (MHz): 1800.00
- L2 Cache (KB): 6815744
- L3 Cache (KB): 12582912
- Total RAM (GB): 15.72

# 7. Insalled Packages

For a list of installed packages, see the `requirements.txt` in the Supplementary Material.

# References

Chen A, Chow A, Davidson A, DCunha A, Ghodsi A, Hong SA, Konwinski A, Mewald C, Murching S, Nykodym T, *et al.* (2020). "Developments in mlflow: A system to accelerate the machine learning lifecycle." In *Proceedings of the fourth international workshop on data management for end-to-end machine learning*, pp. 1–4.

Chen T, Guestrin C (2016). "XGBoost: A Scalable Tree Boosting System." In *Proceedings of the 22nd ACM SIGKDD International Conference on Knowledge Discovery and Data Mining*, KDD '16, pp. 785–794. ACM, New York, NY, USA. ISBN 978-1-4503-4232-2. doi:10.1145/2939672.2939785. URL http://doi.acm.org/10.1145/2939672.2939785.

Chu W, Keerthi SS (2005). "New approaches to support vector ordinal regression." In *Proceedings of the 22nd international conference on Machine learning*, pp. 145–152.

Gutiérrez PA, Perez-Ortiz M, Sanchez-Monedero J, Fernandez-Navarro F, Hervas-Martinez C (2015). "Ordinal regression methods: survey and experimental study." *IEEE Transactions on Knowledge and Data Engineering*, **28**(1), 127–146.

Ke G, Meng Q, Finley T, Wang T, Chen W, Ma W, Ye Q, Liu TY (2017). "Lightgbm: A highly efficient gradient boosting decision tree." *Advances in neural information processing systems*, **30**.

Khwaja A (2012). "KDIGO clinical practice guidelines for acute kidney injury." *Nephron Clinical Practice*, **120**(4), c179–c184.

Liu X, Zou Y, Song Y, Yang C, You J, K Vijaya Kumar B (2018). "Ordinal regression with neuron stick-breaking for medical diagnosis." In *Proceedings of the European Conference on Computer Vision (ECCV) Workshops*, pp. 0–0.

Mahani AS, Sharabiani MT (2017). "Multivariate-From-Univariate MCMC Sampler: The R Package MfUSampler." *Journal of Statistical Software*, **78**, 1–22.

Sun BY, Li J, Wu DD, Zhang XM, Li WB (2009). "Kernel discriminant learning for ordinal regression." *IEEE Transactions on Knowledge and Data Engineering*, **22**(6), 906–910.

Wright SJ (2015). "Coordinate descent algorithms." *Mathematical programming*, **151**(1), 3–34.

Xie JC, Pun CM (2020). "Deep and ordinal ensemble learning for human age estimation from facial images." *IEEE Transactions on Information Forensics and Security*, **15**, 2361–2374.

# A. Detailed Derivations

In this section we derive the key gradients used in the coordinate descent procedure to minimize the negative log-likelihood

$$\mathcal{L}(g, \boldsymbol{\theta}) = -\sum_{n=1}^{N} \ln\Big[\Phi\big(\theta_{y_n+1} - g(\mathbf{x}_n)\big) - \Phi\big(\theta_{y_n} - g(\mathbf{x}_n)\big)\Big],$$

where $\Phi(\cdot)$ denotes the CDF of the chosen link function (e.g., the standard normal CDF for the probit model), and $\phi(\cdot)$ is its corresponding PDF.

## A.1. Derivative with Respect to $g(\mathbf{x})$

For a single observation $(\mathbf{x}_n, y_n)$, define the per-observation loss

$$\ell_n(g, \boldsymbol{\theta}) = -\ln\Big[\Phi\big(\theta_{y_n+1} - g(\mathbf{x}_n)\big) - \Phi\big(\theta_{y_n} - g(\mathbf{x}_n)\big)\Big].$$

Differentiating with respect to $g(\mathbf{x}_n)$ (using the chain rule) yields:

$$\frac{\partial \ell_n}{\partial g(\mathbf{x}_n)} = -\frac{1}{\Phi\big(\theta_{y_n+1} - g(\mathbf{x}_n)\big) - \Phi\big(\theta_{y_n} - g(\mathbf{x}_n)\big)} \left[-\phi\big(\theta_{y_n+1} - g(\mathbf{x}_n)\big) + \phi\big(\theta_{y_n} - g(\mathbf{x}_n)\big)\right].$$

Simplifying, we obtain

$$\frac{\partial \ell_n}{\partial g(\mathbf{x}_n)} = \frac{\phi\big(\theta_{y_n+1} - g(\mathbf{x}_n)\big) - \phi\big(\theta_{y_n} - g(\mathbf{x}_n)\big)}{\Phi\big(\theta_{y_n+1} - g(\mathbf{x}_n)\big) - \Phi\big(\theta_{y_n} - g(\mathbf{x}_n)\big)}.$$

In our boosting algorithm, the negative of this derivative,

$$-\frac{\partial \ell_n}{\partial g(\mathbf{x}_n)},$$

is used as the pseudo-residual for fitting the base learner at each iteration.

### A.2. Derivative with Respect to $\theta_m$

The threshold $\theta_m$ appears in the likelihood terms for those observations whose class label is either $m$ or $m-1$. More precisely, for an observation $n$:

- If $y_n = m-1$, then $\theta_m$ appears as the upper limit in the term $\Phi(\theta_m - g(\mathbf{x}_n))$.

- If $y_n = m$, then $\theta_m$ appears as the lower limit in the term $\Phi(\theta_{m+1} - g(\mathbf{x}_n)) - \Phi(\theta_m - g(\mathbf{x}_n))$.

A unified expression for the derivative with respect to $\theta_m$ is given by

$$\frac{\partial \ell_n}{\partial \theta_m} = -\frac{\phi(\theta_m - g(\mathbf{x}_n))\left[\delta_{y_n,m} - \delta_{y_n,m+1}\right]}{\Phi(\theta_{y_n+1} - g(\mathbf{x}_n)) - \Phi(\theta_{y_n} - g(\mathbf{x}_n))},$$

where $\delta_{i,j}$ denotes the Kronecker delta. This form captures the fact that for observations with $y_n = m$ the threshold contributes positively (as a lower bound), and for those with $y_n = m-1$ it contributes negatively (as an upper bound). In our implementation, the derivative is computed separately for the two groups (denoted $S_m$ and $S_{m+1}$ in the code), and then combined as:

$$\frac{\partial \mathcal{L}}{\partial \theta_m} = -\sum_{n \in S_m} \frac{\phi(\theta_m - g(\mathbf{x}_n))}{D_n} + \sum_{n \in S_{m+1}} \frac{\phi(\theta_m - g(\mathbf{x}_n))}{D_n},$$

with

$$D_n = \Phi(\theta_{y_n+1} - g(\mathbf{x}_n)) - \Phi(\theta_{y_n} - g(\mathbf{x}_n)).$$

A detailed step-by-step algebraic derivation is omitted here for brevity but follows from applying the chain rule to the logarithm of the difference of two cumulative probabilities.

## B. Initialization of $g_0$ and $\boldsymbol{\theta}$

A common initialization for the latent function is to set

$$g_0(\mathbf{x}) \equiv 0.$$

Under this assumption, the predicted probability of class $m$ becomes

$$P(y = m) = \Phi(\theta_{m+1}) - \Phi(\theta_m).$$

Let $\hat{p}_m$ denote the empirical proportion of observations in class $m$. Matching the empirical frequencies to the model probabilities yields

$$\Phi(\theta_{m+1}) - \Phi(\theta_m) \approx \hat{p}_m.$$

By accumulating the class probabilities, one obtains

$$\sum_{j=0}^{m} \hat{p}_j \approx \Phi(\theta_{m+1}).$$

Thus, the initial thresholds are set as

$$\theta_{m+1} = \Phi^{-1}\left(\sum_{j=0}^{m} \hat{p}_j\right), \quad m = 0, \ldots, M-1,$$

with the conventions $\theta_0 = -\infty$ and $\theta_M = +\infty$. This initialization ensures that the model's initial class probabilities are in agreement with the observed empirical distribution.

## C. Learning Rate and Line Search Pseudo-code

Updating the threshold vector $\boldsymbol{\theta}$ is performed via an adaptive line search that seeks a step size $\eta_\theta$ to ensure a decrease in the loss function while maintaining the strict ordering of thresholds. The basic idea is to propose an update

$$\Delta\boldsymbol{\theta} = -\eta_\theta \, \nabla_\theta \mathcal{L},$$

and accept the update if the new thresholds $\boldsymbol{\theta}' = \boldsymbol{\theta} + \Delta\boldsymbol{\theta}$ satisfy:

1. The loss decreases, i.e., $\mathcal{L}(\boldsymbol{\theta}') < \mathcal{L}(\boldsymbol{\theta})$.

2. The thresholds remain strictly ordered: $\theta'_m < \theta'_{m+1}$ for all $m$.

The adaptive line search employs a doubling/halving strategy, which can be summarized as follows:

---
**Algorithm 1** Adaptive Line Search for Threshold Updates
---
1: **Input:** Current thresholds $\boldsymbol{\theta}$, gradient $d\boldsymbol{\theta}$, initial learning rate $\eta_\theta$, factor $\alpha \in (0,1)$ (e.g., 0.5)
2: **Output:** Updated thresholds $\boldsymbol{\theta}'$ and final learning rate $\eta'_\theta$
3: Compute proposed thresholds: $\boldsymbol{\theta}' \leftarrow \boldsymbol{\theta} - \eta_\theta \, d\boldsymbol{\theta}$
4: **if** $\mathcal{L}(\boldsymbol{\theta}') < \mathcal{L}(\boldsymbol{\theta})$ **and** $\boldsymbol{\theta}'$ is strictly ordered **then** ▷ The update is acceptable; try to enlarge the step.
5:     **while** $\mathcal{L}(\boldsymbol{\theta} - \frac{\eta_\theta}{\alpha} \, d\boldsymbol{\theta}) < \mathcal{L}(\boldsymbol{\theta})$ **and** ordering is maintained **do**
6:         Update: $\eta_\theta \leftarrow \eta_\theta/\alpha$
7:         $\boldsymbol{\theta}' \leftarrow \boldsymbol{\theta} - \eta_\theta \, d\boldsymbol{\theta}$
8:     **end while**
9: **else** ▷ The update is not acceptable; reduce the step size.
10:     **while** $\mathcal{L}(\boldsymbol{\theta} - \eta_\theta \, d\boldsymbol{\theta}) \geq \mathcal{L}(\boldsymbol{\theta})$ **or** ordering is violated **do**
11:         Update: $\eta_\theta \leftarrow \alpha \, \eta_\theta$
12:         $\boldsymbol{\theta}' \leftarrow \boldsymbol{\theta} - \eta_\theta \, d\boldsymbol{\theta}$
13:     **end while**
14: **end if**
15: **Return:** $\boldsymbol{\theta}'$ and $\eta_\theta$
---

This algorithm ensures that the chosen step size is as large as possible while still guaranteeing a decrease in the loss and maintaining the monotonicity of the thresholds.

**Affiliation:**

Mansour T.A. Sharabiani  
Imperial College London  
London, UK  
E-mail: mt5605@imperial.ac.uk  
Alex Bottle  
Imperial College London  
London, UK  
E-mail: robert.bottle@imperial.ac.uk  
Alireza S. Mahani  
Statman Solution Ltd.  
London, UK  
E-mail: alireza.s.mahani@gmail.com